\documentstyle[12pt,epsfig,rotating] {article}
\begin{document}

\newcommand{\Z}{\mbox{$\mathrm{Z}$}}
\newcommand{\Zo}{\mbox{$\mathrm{Z^0}$}}
\newcommand{\W}{\mbox{$\mathrm{W}$}}
\newcommand{\bZo}{{\bf \mbox{$\mathrm{Z}$}}}
\newcommand{\Zg}{\mbox{$\mathrm{Z}^{0}\gamma$}}
\newcommand{\ZZ}{\mbox{$\mathrm{Z}^{0}\mathrm{Z}^{0}$}}
\newcommand{\WW}{\mbox{$\mathrm{W}^+\mathrm{W}^-$}}
\newcommand{\Zs}{\mbox{$\mathrm{Z}^{*}$}}
\newcommand{\h}{\mbox{$\mathrm{h}^{0}$}}
\newcommand{\Ho}{\mbox{$\mathrm{H}$}}
\newcommand{\ho}{\mbox{$\mathrm{h}$}}
\newcommand{\Hp}{\mbox{$\mathrm{H}^{+}$}}
\newcommand{\Hm}{\mbox{$\mathrm{H}^{-}$}}
\newcommand{\Hsm}{\mbox{$\mathrm{H}^{0}_{SM}$}}
\newcommand{\A}{\mbox{$\mathrm{A}^{0}$}}
\newcommand{\Hpm}{\mbox{$\mathrm{H}^{\pm}$}}
\newcommand{\X}{\mbox{${\tilde{\chi}^0}$}}
\newcommand{\ko}{\mbox{${\tilde{\chi}^0}$}}
\newcommand{\ee}{\mbox{$\mathrm{e}^{+}\mathrm{e}^{-}$}}
\newcommand{\bee}{\mbox{$\boldmath {\mathrm{e}^{+}\mathrm{e}^{-}} $}}
\newcommand{\mm}{\mbox{$\mu^{+}\mu^{-}$}}
\newcommand{\nn}{\mbox{$\nu \bar{\nu}$}}
\newcommand{\qq}{\mbox{$\mathrm{q} \bar{\mathrm{q}}$}}
\newcommand{\pb}{\mbox{$\mathrm{pb}^{-1}$}}
\newcommand{\ra}{\mbox{$\rightarrow$}}
\newcommand{\br}{\mbox{$\boldmath {\rightarrow}$}}
\newcommand{\tautau}{\mbox{$\tau^{+}\tau^{-}$}}
\newcommand{\ga}{\mbox{$\gamma$}}
\newcommand{\gamgam}{\mbox{$\gamma\gamma$}}
\newcommand{\tp}{\mbox{$\tau^+$}}
\newcommand{\tm}{\mbox{$\tau^-$}}
\newcommand{\tpm}{\mbox{$\tau^{\pm}$}}
\newcommand{\uu}{\mbox{$\mathrm{u} \bar{\mathrm{u}}$}}
\newcommand{\dd}{\mbox{$\mathrm{d} \bar{\mathrm{d}}$}}
\newcommand{\bb}{\mbox{$\mathrm{b} \bar{\mathrm{b}}$}}
\newcommand{\cc}{\mbox{$\mathrm{c} \bar{\mathrm{c}}$}}
\newcommand{\mumu}{\mbox{$\mu^+\mu^-$}}
\newcommand{\csbar}{\mbox{$\mathrm{c} \bar{\mathrm{s}}$}}
\newcommand{\cbars}{\mbox{$\bar{\mathrm{c}}\mathrm{s}$}}
\newcommand{\nunu}{\mbox{$\nu \bar{\nu}$}}
\newcommand{\nubar}{\mbox{$\bar{\nu}$}}
\newcommand{\mQ}{\mbox{$m_{\mathrm{Q}}$}}
\newcommand{\mZ}{\mbox{$m_{\mathrm{Z}}$}}
\newcommand{\mH}{\mbox{$m_{\mathrm{H}}$}}
\newcommand{\mHp}{\mbox{$m_{\mathrm{H}^+}$}}
\newcommand{\mh}{\mbox{$m_{\mathrm{h}}$}}
\newcommand{\mA}{\mbox{$m_{\mathrm{A}}$}}
\newcommand{\mHpm}{\mbox{$m_{\mathrm{H}^{\pm}}$}}
\newcommand{\mHsm}{\mbox{$m_{\mathrm{H}^0_{SM}}$}}
\newcommand{\mW}{\mbox{$m_{\mathrm{W}^{\pm}}$}}
\newcommand{\mt}{\mbox{$m_{\mathrm{t}}$}}
\newcommand{\mb}{\mbox{$m_{\mathrm{b}}$}}
\newcommand{\lpm}{\mbox{$\ell ^+ \ell^-$}}
\newcommand{\G}{\mbox{$\mathrm{GeV}$}}
\newcommand{\Gc}{\mbox{${\rm GeV}/c$}}
\newcommand{\Gcs}{\mbox{${\rm GeV}/c^2$}}
\newcommand{\Mcs}{\mbox{${\rm MeV}/c^2$}}
\newcommand{\sba}{\mbox{$\sin ^2 (\beta -\alpha)$}}
\newcommand{\cba}{\mbox{$\cos ^2 (\beta -\alpha)$}}
\newcommand{\tanb}{\mbox{$\tan \beta$}}
\newcommand{\sqrts}{\mbox{$\sqrt {s}$}}
\newcommand{\sqrtsp}{\mbox{$\sqrt {s'}$}}
\newcommand{\msusy}{\mbox{$M_{\rm SUSY}$}}
\newcommand{\mg}{\mbox{$m_{\tilde{\rm g}}$}}
\begin{titlepage}
\begin{center}
\vspace{-0.5cm}
{\Large EUROPEAN ORGANIZATION FOR NUCLEAR RESEARCH}
\end{center}
%\bigskip
\begin{flushright}
  LHWG note 2001-05 \\
  ALEPH  2001-043 PHYSICS 2001-016 \\
  DELPHI 2001-115 CONF 538 \\
  L3 Note 2689 \\
  OPAL Technical Note TN696\\
  4 July, 2001 \\
%  {\today} 
\end{flushright}
%\bigskip
\begin{center}{\Large \bf Search for Charged Higgs bosons: \\
Preliminary Combined Results Using LEP data \\ 
Collected at Energies up to 209~GeV}
 \bigskip
\end{center}
\begin{center}
      {\large  ALEPH, DELPHI, L3 and OPAL Collaborations}\\
      {\large The LEP working group for Higgs boson searches}
%\footnote{      Contributions from
%       P.~Colas, P.~Garcia-Abia, C.~Hajdu, K.~Hoffman,
%       D.~Horv\'ath, P.~Igo-Kemenes, A.~Kiiskinen, M.C.~Lemaire, 
%       E.~Locci, P.~Lutz, F.~Matorras, A.N.~Okpara, C.~Tully.}
\end{center}
\bigskip
\begin{center}{\Large  Abstract}\end{center}
In the year 2000 the four LEP experiments have collected 
870~\pb\ of data at energies between 200 and 209 GeV, 
with about 510~\pb\ above 206~GeV. These data have been combined with
data sets collected earlier at lower energies. 
%The following 95\% confidence level bounds have been obtained. 
For charged Higgs bosons predicted by two-doublet extensions of the Standard 
Model and decaying only into the channels \Hp\ra\csbar\ and \tp$\nu$, 
%no statistically significant excess has been observed when 
%compared to the Standard Model background prediction, and 
a lower mass bound of 78.6~\Gcs\ is obtained, 
at the 95\% confidence level. 

\begin{center}
ALL RESULTS QUOTED IN THIS NOTE ARE PRELIMINARY\\
(contributed paper for EPS'01 in Budapest and LP'01 in Rome)
\end{center}
%
%\bigskip
\end{titlepage}
\section{Introduction}
We present combined results from the ALEPH, DELPHI, L3 and OPAL 
Collaborations on searches for the charged Higgs boson
predicted by extensions of the Standard Model with two 
Higgs field doublets. The results are obtained by   
combining the data collected in the year 2000 at centre-of-mass 
energies between 200 and 209~GeV with data collected earlier at
energies of 189~GeV and larger~\cite{adlo-cernep}. The total luminosity
used in this combination is 2500~\pb\. The new data represent an
integrated luminosity of approximately 870~\pb\ in total, with
about 510~\pb\ above 206~GeV.

The cross-sections, branching ratios and other physics quantities  
used in this combination of data are calculated within the 
HZHA program package~\cite{hzha}. 

Each experiment has generated Monte Carlo event samples for 
the Higgs signal and the various background processes, typically
at 202, 204, 206, 208 and 210~GeV energies. Cross-sections, 
branching ratios, distributions of the reconstructed mass and 
other discriminating variables relevant to the combination have 
been interpolated to the energies which correspond to the data sets. 
%In this procedure special care has been taken 
%to the regions of kinematic cutoff where the signal and
%background distributions vary rapidly. 
It has been checked that the interpolation procedures do not add 
significantly to the final systematic errors.

The statistical procedure adopted for the combination of the data 
and the precise definitions of the confidence levels 
$CL_b,~CL_{s+b},~CL_s$ by which the search results are expressed, 
have been previously described~\cite{adlo-cernep}~\cite{adlo-earlier}.
The main sources of systematic error affecting the signal and 
background predictions are included. This is done using an
extension of the method of Cousins and Highland~\cite{cousins-highland} 
where the confidence levels are the averages of a large ensemble of 
Monte Carlo experiments. The correlations between 
search channels, LEP energies and individual experiments have not been
taken into account, but these correlations are estimated to have
only small effects to the final results.
%each one with a different choice of
%signal and background, varied within the errors.

%%%%%%%%%%%%%%%%%%% CHARGED HIGGS  %%%%%%%%%%%%%%%%%%%%%%%%%%%%%%%%%%%%%%%%%%%%%%
\section{Combined searches for the charged Higgs boson}
Charged Higgs bosons are predicted by models with two 
Higgs field doublets (2HD models), of which the MSSM is a particular case
with supersymmetry. 
At LEP2 energies charged Higgs bosons are expected to be
produced mainly through the process \ee\ra\Hp\Hm. In the MSSM, at tree-level,
the \Hpm\ is constrained to be heavier than the W$^{\pm}$ bosons, but for
specific choices of the MSSM parameters,
loop corrections can drive the mass to lower values. Thus,
%sensitivity of current searches is limited to the range below \mW\ due to the 
%background from \ee\ra~W$^+$W$^-$,  
any signal for \Hp\Hm\ would
indicate either new physics beyond the MSSM or a rather extreme set of MSSM 
parameter values.

In 2HD models the \Hpm\ mass is not predicted, and the tree-level 
cross-section is fully determined by the mass~\cite{djouadi}.
The searches are carried out under the assumption that 
the two decays \Hp\ra\csbar\ and \Hp\ra\tp$\nu$ exhaust the \Hp\ decay
width, but the relative branching ratio is free. Thus, the 
searches encompass the following \Hp\Hm\ final states: (\csbar)(\cbars),
(\tp$\nu$)(\tm\nubar) and the mixed mode (\csbar)(\tm\nubar)+(\cbars)(\tp$\nu$).
The combined search results are presented as a function of the branching 
ratio B(\Hp\ra\tp$\nu$).

\begin{figure}[htb]
\begin{center}
{\large CHARGED HIGGS - PRELIMINARY} \\
\epsfig{figure=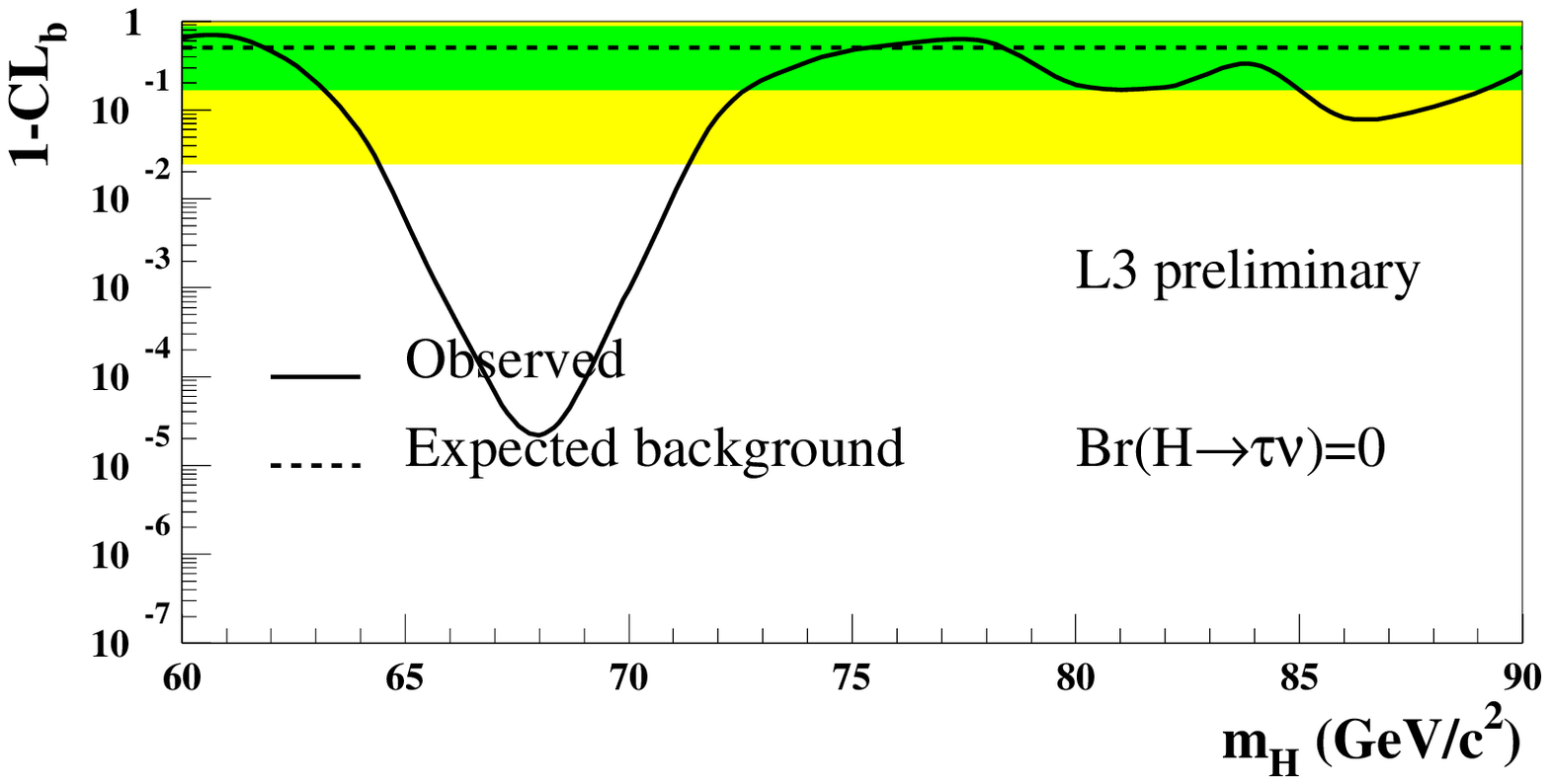,width=0.85\textwidth}
\epsfig{figure=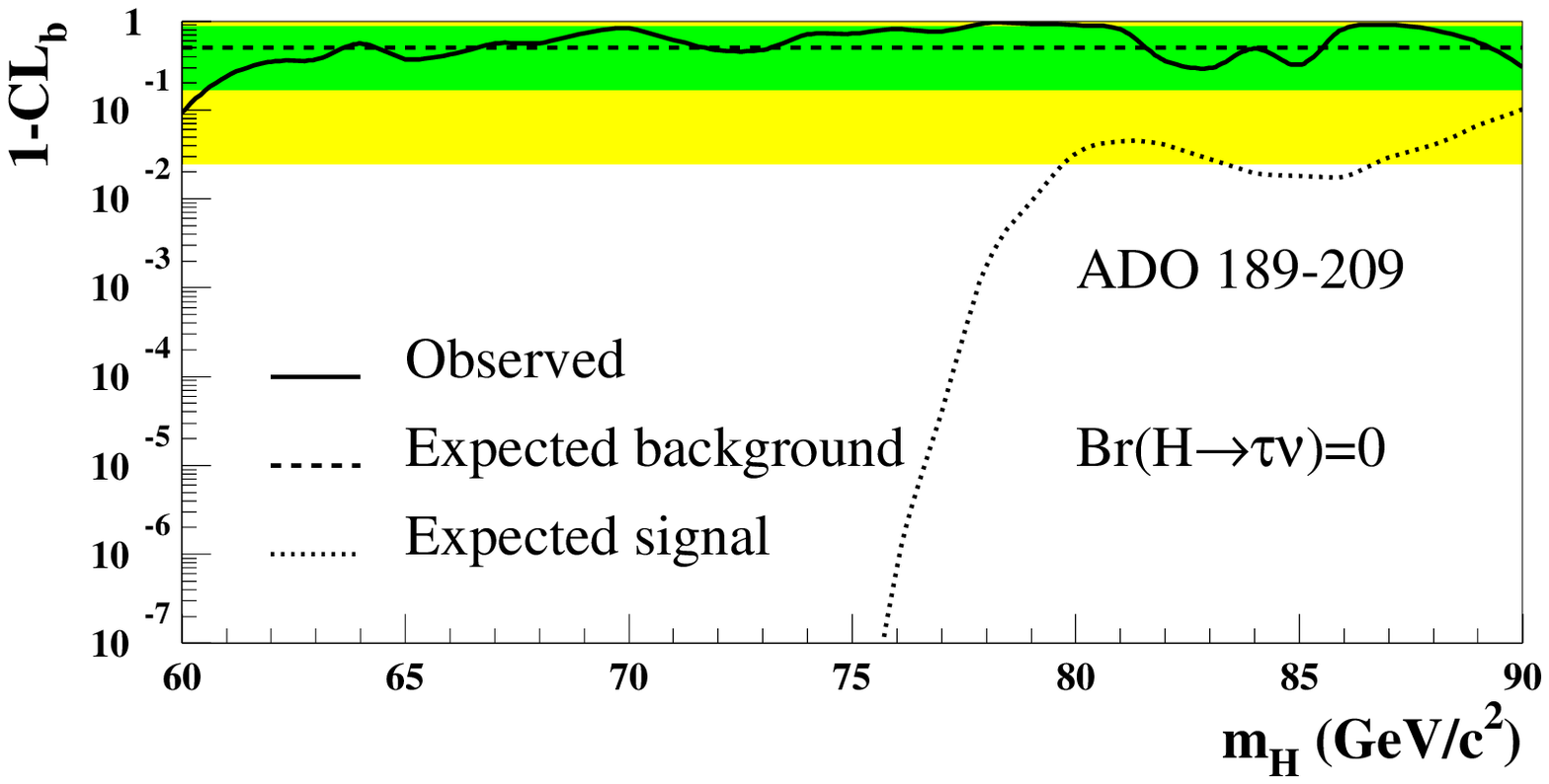,width=0.85\textwidth}
\caption[]{\small \it 
The confidence level $1- CL_b$  as a function 
of \mHpm, for the branching ratio
B(\Hp\ra\tp$\nu$)=0 (top : L3 alone, bottom : ADO combination).  
The straight horizontal line at 50\% and the shaded
bands represent the mean result and the symmetric $1\sigma$ and 
$2\sigma$ probability bands expected in the absence of a
signal. The solid curve is the observed result and 
the dotted curve shows the median result expected for a
signal when tested at the ``true" mass.
\label{ado}}
\end{center}
\end{figure}

Details of the searches of the individual experiments 
%using the data collected at energies between 200 and 209~GeV (year 2000 data),
can be found in~\cite{charged-inputs}\cite{l3note}. 
These are summarised in Table~\ref{table-ch-input}, 
together with the 95\% CL lower bounds, expected and observed. 
In the current combination, the OPAL search in the mixed channel is not
included since it is still under scrutiny.
In the table we quote 
the mass limits separately for B(\Hp\ra\tp$\nu$) = 0, 1, 
and a limit independent of the branching ratio. It should be noted that
L3 observes an excess of events in the pure hadronic and the semi-leptonic
channels in the mass region around 68~GeV~\cite{l3note}.
This behaviour is reproduced when the combination protocoll is applied
to the L3 data only, but is not seen when applied to the ALEPH, 
DELPHI and OPAL data, see Figure~\ref{ado}. 
Indeed, ALEPH data show a clear deficit around that mass. 
The compatibiliy of the L3 and the three other experiments observations
in the vicinity of 68~GeV is under investigation.
%Plots of the candidate masses can be seen in Figure~\ref{mass_charged} 
%for the hadronic and semileptonic final states.

%\clearpage

\begin{figure}[htb]
\begin{center}
{\large CHARGED HIGGS - PRELIMINARY} \\
\epsfig{figure=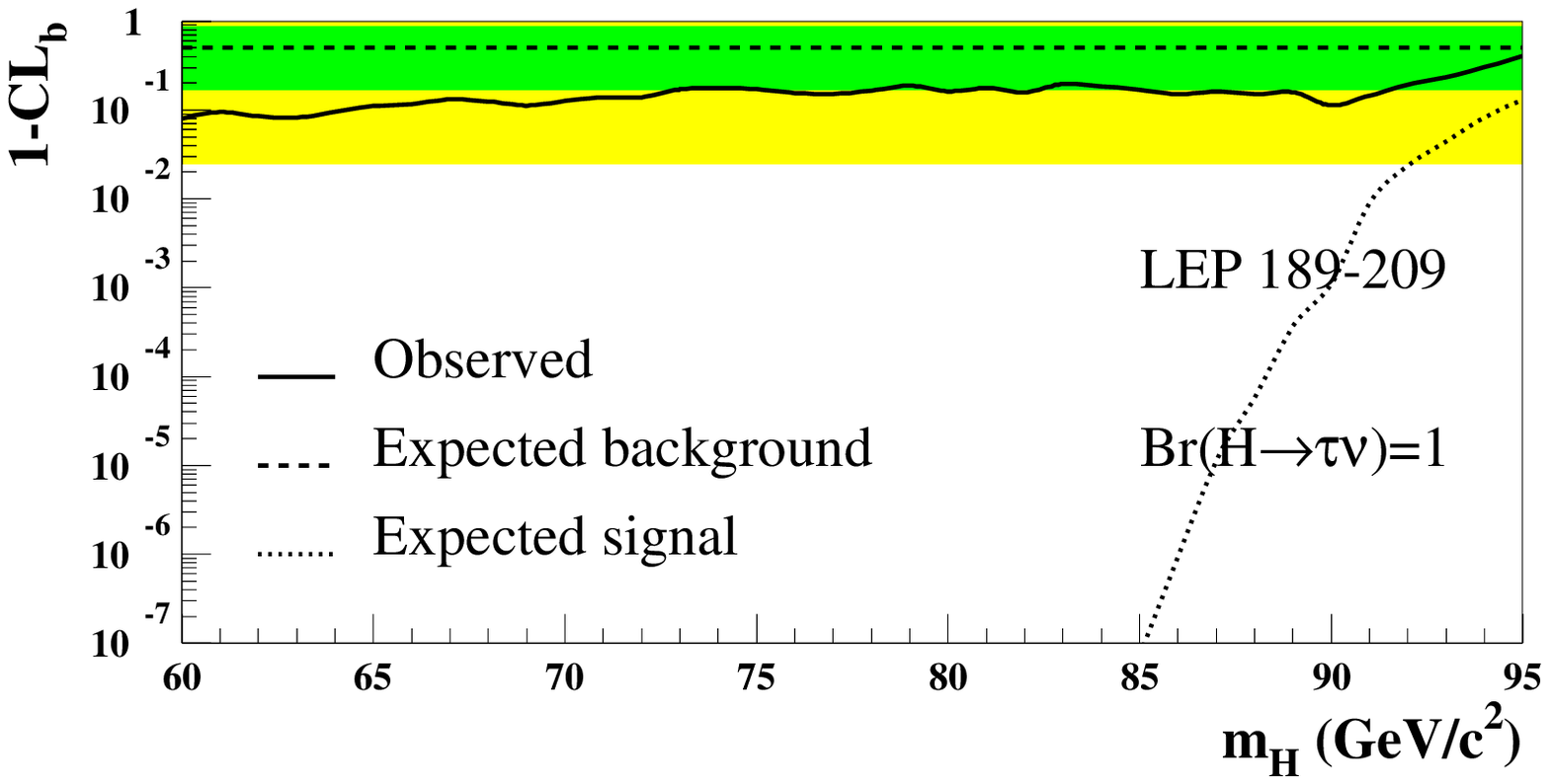,width=0.85\textwidth}
\epsfig{figure=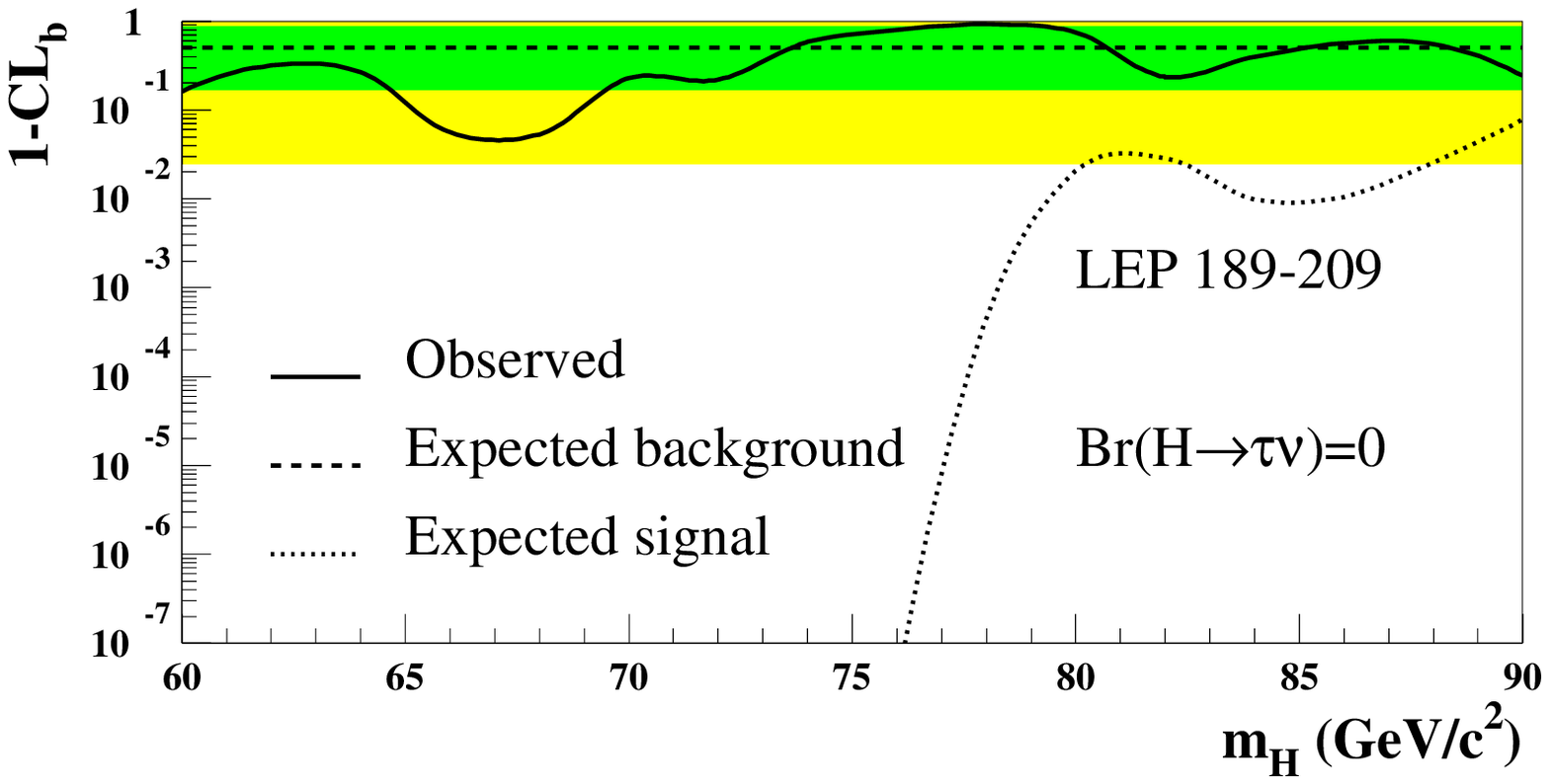,width=0.85\textwidth}
\caption[]{\small \it 
The confidence level $1- CL_b$  as a function 
of \mHpm, for the branching ratio
B(\Hp\ra\tp$\nu$)=0 and 1 (separate plots).  
The straight horizontal line at 50\% and the shaded
bands represent the mean result and the symmetric $1\sigma$ and 
$2\sigma$ probability bands expected in the absence of a
signal. The solid curve is the observed result and 
the dotted curve shows the median result expected for a
signal when tested at the ``true" mass.
\label{charged-clb}}
\end{center}
\end{figure}

%In order to search for a possible signal, the test mass \mHpm\ has been 
%scanned.
%The test-statistic $X=-2ln(Q)=\Delta\chi^2$ versus the test mass 
%\mHpm\ is shown in Figure~\ref{ch-lnq}, 
%separately for the branching ratio B(\Hp\ra\tp$\nu$) fixed to 1, 0.5 and 0. 
 Assuming that a statistical combination is justified,
% Combining the results from the four experiments,
 Figure~\ref{charged-clb} shows the background confidence level 
$1-CL_b$ as a function of \mHpm, expected and observed, for 
B(\Hp\ra\tp$\nu$)=1 and 0. 
The observed confidence level is everywhere within the light-shaded  
$\pm 2\sigma$ bands of the background prediction.
%except 
%... *** Some discussion ... ***   proposal
%for some low mass regions, from 60 to 67~\Gcs\  at B(\Hp\ra\tp$\nu$)=0.5 and
%around 66~\Gcs\ at B(\Hp\ra\tp$\nu$)=1 that never reach the 3 $\sigma$ level
%and are far from the expectation for a signal.
%These small anomalies are under investigation, which are not the result
%of a tendency shared by the 4 experiments.

\begin{figure}[htb]
\begin{center}
{\large CHARGED HIGGS - PRELIMINARY} \\
\epsfig{figure=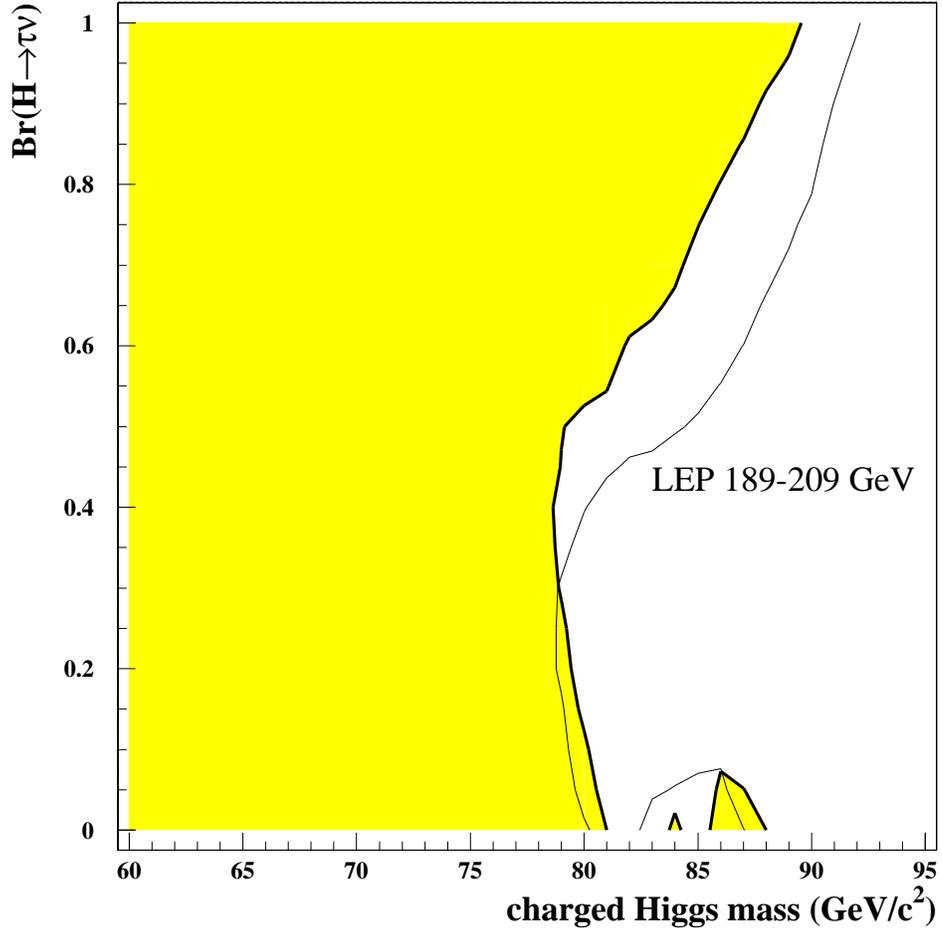,width=14cm}
\caption[]{\small \it 
The 95\% CL bounds on \mHpm\ as a function of the branching ratio
B(\Hp\ra\tp$\nu$), combining the data collected by the four LEP experiments at 
energies from 189 to 209~GeV. 
The expected exclusion limits are indicated by the thin solid line and the
observed limits by the thick solid one. The shaded area is excluded at the 
95\% CL. 
%The light full lines show the observed limits channel by channel. 
\label{charged-limit}}
\end{center}
\end{figure}

The mass limits expected and observed are shown in 
Figure~\ref{charged-limit}. To obtain the limits, 
the branching ratio B(\Hp\ra\tp$\nu$) has been scanned in steps of 
0.05, and the limit setting procedure repeated for each step. 
In the hadronic channel and for masses close to 
\mW, the sensitivity is suppressed by the large \ee\ra\WW\ background.
There is a regain  of sensitivity at higher masses, as signalled
by the excluded ``islands'' above 84 \Gcs.

The combined 95\% CL bounds are listed in 
Table~\ref{table-ch-limits} for B(\Hp\ra\tp$\nu$)=0, 1, and for the 
weakest limit, valid for any value of B(\Hp\ra\tp$\nu$).
Taking the lowest 
of the observed limits from Table~\ref{table-ch-limits},
we quote a 95\% CL lower bound of 78.6~\Gcs\  for the
mass of the charged Higgs boson.  
The inclusion of systematic errors has shifted 
the observed mass limits downwards by 600, 600, and 200~\Mcs\ 
for B(\Hp\ra\tp$\nu$)=0, 0.5 and 1, respectively. A major source
of systematic error comes from the measurement of the W mass, since
a shift of 50~\Mcs\ in that value induces a shift of 200~\Mcs\ in the
limits when these limits are around the W mass.

%Recent calculations~\cite{racoon} predict a cross-section for W$^+$W$^-$
%production
%which is lower by a few percent than the ones provided by earlier event
%generators~\cite{gentle}. 
%Had one chosen the new calculations for the estimation of the background
%would have decreased
%the mass limit by ***, ***, and ***~\Mcs\ for B(\tp$\nu$)=0, 1, and for
%arbitrary value,
%respectively.

As a cross-check of the confidence level calculation procedures, 
the expected and observed limits have been 
cross-checked using another test-statistic 
(Method C in~\cite{adlo-earlier}). %Similar results were found.
%The limits were within $\pm 700$~\Mcs\ of the quoted values.

\begin{figure}[htb]
\begin{center}
{\large CHARGED HIGGS -PRELIMINARY} \\
\epsfig{figure=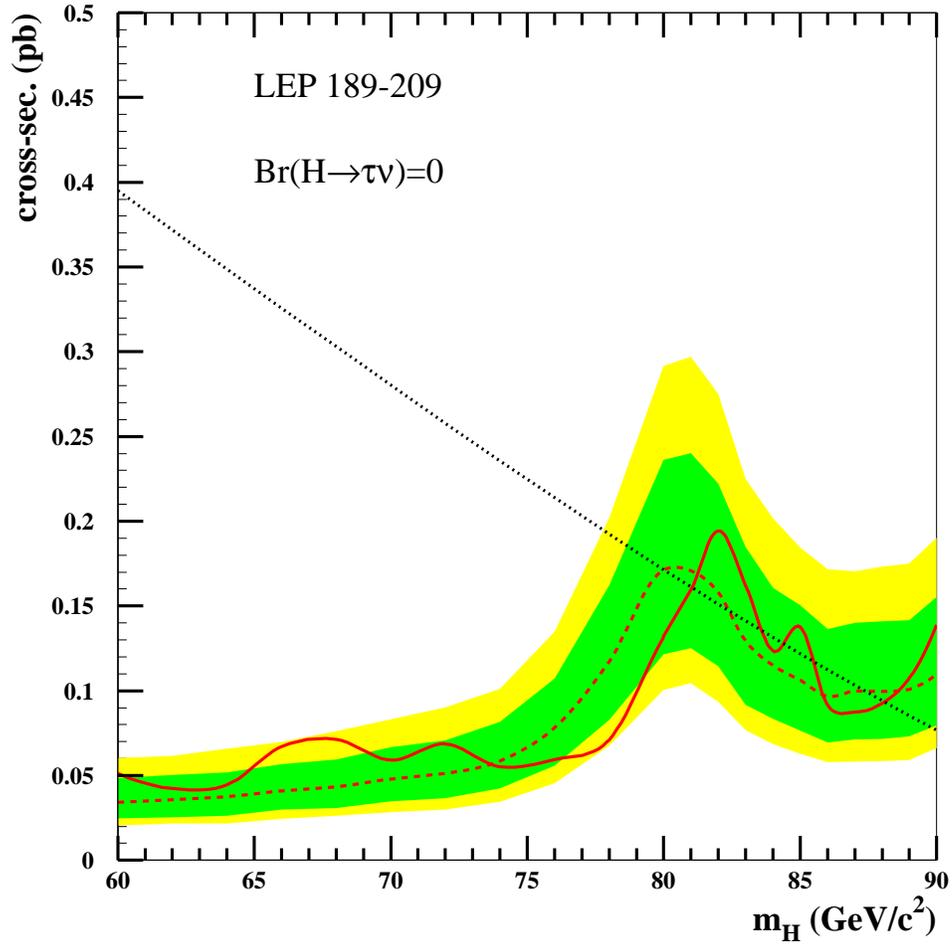,width=14cm}
\caption[]{\small \it 
The 95\% CL bound on the production cross-section as a function of 
\mHpm\  for a branching ratio B(\Hp\ra\tp$\nu$)=0, 
combining the data collected by the four LEP experiments at 
energies from 189 to 209~GeV. 
The expected exclusion limits are indicated by the dashed line 
and the shaded bands (1 and 2 $\sigma$), the observed limit 
by the solid line. The dotted line represents the 2HD model 
computed charged Higgs cross-section at 206 GeV.
\label{xsec-limit}}
\end{center}
\end{figure}

Figure~\ref{xsec-limit} shows the 95\% CL upper bound on 
the cross-section
(with $\pm$ 1$\sigma$ and $\pm$ 2$\sigma$ bands) for the hadronic decay 
topology \Hp\ra\csbar\ . The dotted line corresponds to the 2HD model 
prediction at 206 GeV. Its intersections with the observed bound 
(full line) define the excluded regions in Figure~\ref{charged-limit} 
for B(\Hp\ra\tp$\nu$)=0. The relation between the features 
of Figure~\ref{charged-clb}(lower part) and  
Figure~\ref{xsec-limit} is apparent.
%Another way to summarize the results of this combination at a given 
%B(\Hp\ra\tp$\nu$) is presented on Figure~\ref{xsec-limit} in the case 
%B(\Hp\ra\tp$\nu$)=0. It shows
%the production cross-section which can be excluded at the 95\% CL as a
%function of the test mass. Thus, the intersections with the thick black
%curve showing our present computation of the charged Higgs 
%cross-section give directly the limits (observed and/or expected) 
%already shown and explain where the ``islands'' come from, while the 
%behaviour of the observed excluded
%cross-section inside the 2$\sigma$ bands of the expectation in case of
%background only is a direct reflect of what was seen on the (1-CL$_b$)
%curve.

% 
\begin{table} [hbtp]
\begin{center}
\begin{tabular}{||l||c|c|c|c||}
\hline\hline
Experiment:                             & ALEPH    & DELPHI   & L3        & 
OPAL      \\
\hline
%$<204.5$~GeV: Int. luminosity (\pb):    &          &          &           &           \\
%\phantom{.....}Backg. exp. / Events obs.                                    &&&&      \\
%\phantom{..........}(\csbar)(\cbars) :  &          &          &           &           \\
%\phantom{..........}(\csbar)(\tp$\nu$): &          &          &           &           \\
%\phantom{..........}(\tp$\nu$)(\tm\nubar):&        &          &           &           \\
%\hline
%204.5-205.5~GeV: Int. luminosity (\pb): &          &          &           &           \\
%\phantom{.....}Backg. exp. / Events obs.                                    &&&&\\
%\phantom{..........}(\csbar)(\cbars) :  &          &          &           &           \\
%\phantom{..........}(\csbar)(\tp$\nu$): &          &          &           &           \\
%\phantom{..........}(\tp$\nu$)(\tm\nubar):&        &          &           &           \\
%\hline
%$>$205.5~GeV: Int. luminosity (\pb):    &          &          &           &           \\
%\phantom{.....}Backg. exp. / Events obs.                                    &&&&\\
%\phantom{..........}(\csbar)(\cbars) :  &          &          &           &           \\
%\phantom{..........}(\csbar)(\tp$\nu$): &          &          &           &           \\
%\phantom{..........}(\tp$\nu$)(\tm\nubar):&        &          &           &           \\
%\hline\hline
Total: Int. luminosity (\pb):           &217.2     & 225.1    & 217.8  &217.4      \\
\phantom{.....}Backg. exp. / Events obs. (*)             &&&&\\
\phantom{..........}(\csbar)(\cbars) :  &997.7/968&412.8/387&883.3/961&424.2/439 \\
\phantom{..........}(\csbar)(\tp$\nu$): &118.0/127&190.8/173&171.8/171&203.5/224 \\
\phantom{..........}(\tp$\nu$)(\tm\nubar):&22.0/17& 23.8/ 25& 49.8/44 &331.7/315\\
\hline\hline
Events in all channels:    &1137.7/1112 &627.4/585 &1104.9/1176 &959.4/978 \\
\hline
Limit exp.(median)/ observed                           &&&& \\
\phantom{.....}for B=0:       &78.1/80.7 & 77.0/77.4&76.5/67.7  &77.1/76.2  \\
\phantom{.....}for B=1:       &86.9/83.4 & 89.3/85.4&84.7/82.8  &86.5/84.5  \\
\phantom{.....}for any B:     &76.9/78.0 & 75.4/73.8&75.1/65.6  &74.5/72.2  \\
\hline\hline
\end{tabular}
\end{center}
\caption{\small\it Individual search results for the
\ee\ra\Hp\Hm final states. The luminosities and numbers of events correspond 
to the data sets taken at energies between 200 and 209~GeV (year 2000 data).
(*) The OPAL selection  is mass-dependent; 
the numbers are given here for \mHpm = 80~\Gcs.
%Also, the OPAL systematic uncertainty is handled conservatively
%by reducing the subtractable background by 1 standard deviation.   
\label{table-ch-input}}
\end{table}

\begin{table} [hbtp]
\begin{center}
\begin{tabular}{||l||c||}
\hline\hline
                 & Mass limit in \Gcs\ (95\% CL) \\
\hline
B(\Hp\ra\tp$\nu$)=0                  & \\
Limit expected (median) : & 80.2 (*)   \\
Limit observed :          & 81.0 (*)   \\
\hline
B(\Hp\ra\tp$\nu$)=1                  & \\
Limit expected (median) : & 92.1  \\
Limit observed :          & 89.6  \\
\hline
Any B(\Hp\ra\tp$\nu$)                & \\
Limit expected (median):  & 78.8  \\
Limit observed :          & 78.6  \\
\hline\hline
\end{tabular}
\end{center}
\caption{\small\it The combined 95\% CL lower bounds for the mass of the
charged Higgs boson, expected and observed, for fixed 
values of the branching ratio B(\Hp\ra\tp$\nu$)
and for the B(\Hp\ra\tp$\nu$) giving the weakest limit. 
(*) These limits do not take into account the small regions excluded 
above 82 GeV.
\label{table-ch-limits}}
\end{table}
%

%%%%%%%%%%%%%%%%%%%%%%%%%%%%%%%%%%%%%%%%%%%%%%%%%%%%%%%%%%%%%%%%%%%%%%%%%%%%%%%%%%%%%%%%%%%%%%%%%%%
%
%\begin{figure}[htb]
%\begin{center}
%{\large CHARGED HIGGS - PRELIMINARY} \\
%\epsfig{figure=lep_cscs_all.eps,width=0.4\textwidth}
%\epsfig{figure=lep_cstn_all.eps,width=0.4\textwidth}
%\caption[]{\small \it 
%Reconstructed mass for candidate events in the charged Higgs
%searches by the 4 LEP experiments..
%\label{mass_charged}}
%\end{center}
%\end{figure}
%

%\begin{figure}[htb]
%\begin{center}
%{\large CHARGED HIGGS - PRELIMINARY} \\
%\epsfig{figure=L3.eps,width=0.75\textwidth}
%\caption[]{\small \it 
%The background confidence level $1- CL_b$  as a function 
%of \mHpm, for the branching ratio
%B(\Hp\ra\tp$\nu$)=0.1 in L3 (taken from their note).  
%The solid line shows the values computed from the observed results
%and the dashed line the expectation for the background only hypothesis.
%The dotted line is the curve expected for a 68 GeV Higgs signal at
%this value of the branching ratio. The shaded areas represent the symmetric
%1$\sigma$ and 2$\sigma$ probability bands expected in the absence of a
%signal.
%\label{L3}}
%\end{center}
%\end{figure}
%\clearpage
%
\clearpage
%
%\begin{figure}[htb]
%\begin{center}
%{\large CHARGED HIGGS -PRELIMINARY} \\
%\epsfig{figure=adlo_k2_100.eps,width=0.85\textwidth}
%\epsfig{figure=adlo_k2_050.eps,width=0.85\textwidth}
%\epsfig{figure=adlo_k2_000.eps,width=0.85\textwidth}
%\caption[]{\small \it 
%The $\Delta \chi^2$ as a function of \mHpm,
%separately for B(\tp$\nu$)=0, 0.5 and 1. 
%In each case, the dotted line shows 
%the expectation for the background-only hypothesis and the full line the values computed from 
%the observed results.
%The shaded areas show the symmetric $1\sigma$ and $2\sigma$ probability bands for the background hypothesis. 
%\label{ch-lnq}}
%\end{center}
%\end{figure}
%\clearpage
%
\clearpage
\clearpage

%%%%%%%%%%%%%%%%%%%% END CHARGED HIGGS  %%%%%%%%%%%%%%%%%%%%%%%%%%%%%%%%%%%%%%%%%%%%%%%%%

\clearpage
\section{Summary}

The searches of the four LEP experiments for charged Higgs bosons predicted 
by models with two Higgs field doublets were combined. These 
assume that the two decays \Hp\ra\csbar\ and \Hp\ra\tp$\nu$ exhaust 
the \Hp\ decay width. In the absence of a signal, mass 
limits are obtained as a function of the 
branching ratio B(\Hp\ra$\tau^+\nu$). The most general lower limit,
valid at the 95\% confidence level for any value of the branching ratio, is 
78.6~\Gcs. 

\begin{center}
ALL THE RESULTS QUOTED IN THIS NOTE ARE PRELIMINARY.
\end{center}
%\newpage

%%%%%%%%%%%%%%%%%%%%%%%%%%%%%%%%%%%%%%%%%%%%%%%%%%%%%%%%%%%%%%%%%%%%%%%%%%%%%


\begin{thebibliography}{99} 
%
%1
\bibitem{adlo-cernep}
ALEPH, DELPHI, L3 and OPAL Collab., The LEP working group for Higgs boson searches, 
{\it Searches for Higgs bosons: Preliminary combined results using LEP data collected at energies up to 202 GeV},
CERN-EP/2000-055.
%1_bis to take into account UPDATES since Moriond (with BIG impact on the
%our 99 results => our MORIOND note should NOT be quoted any more for our
%'99 inputs)
%
%2
\bibitem{hzha}
HZHA: P. Janot, in CERN Report 96-01, Vol. 2, p. 309 (1996);
Version 3, released in December 1999, http://alephwww.cern.ch/janot/Generators.html.
%
%9
\bibitem{adlo-earlier}
ALEPH, DELPHI, L3 and OPAL Collab., The LEP working group for Higgs boson searches,
CERN EP 98-046 (1998).
%
%3
\bibitem{cousins-highland} R.D.~Cousins and V.L.~Highland, Nucl. Instr. Methods
{\bf A320} (1992) 331. 
%
\bibitem{djouadi}
A. Djouadi, J. Kalinowski and P.M. Zerwas, Z. Phys. {\bf C57} (1993) 569.
%10
\bibitem{charged-inputs} Charged Higgs inputs of the experiments.\\
ALEPH Collab., ALEPH 2001-016 CONF 2001-013 (2001);  \\
%See Reference~\cite{summary-aleph} for the 2000 data update.
DELPHI Collab., DELPHI 2001-030 CONF 471 (2001); \\
OPAL Collab. OPAL PN472 (2001).
\bibitem{l3note}
L3 Collab., Physics Letters {\bf B496} (2000) 34. \\
L3 Collab., L3 note 2643 (2001).
\end{thebibliography}
\end{document}